\documentclass[conference,letterpaper]{IEEEtran}

\usepackage[utf8]{inputenc} 
\usepackage[T1]{fontenc}
\usepackage{url}
\usepackage{ifthen}
\usepackage{cite}
\usepackage[cmex10]{amsmath} 
\usepackage{amsmath}  
\usepackage{amssymb}  
\usepackage{mathrsfs} 
\usepackage{cite}     
\usepackage{comment}  
\usepackage{enumerate}
\usepackage{graphicx}
\usepackage{setspace}
\usepackage{amsthm}
\usepackage{multirow}
\usepackage{color}
\usepackage{longtable}
\usepackage{array}

\usepackage{stmaryrd}

\usepackage{tikz}
\usepackage[export]{adjustbox}

\usetikzlibrary{arrows, arrows.meta,calc,positioning,shapes.arrows}
\usetikzlibrary{decorations.pathreplacing}
\usetikzlibrary{chains,fit,shapes}
\usetikzlibrary {matrix}

\theoremstyle{plain}
\newtheorem{theorem}{Theorem}

\newtheorem{proposition}[theorem]{Proposition}

\theoremstyle{definition}

\newtheorem{example}[theorem]{Example}

\DeclareMathAlphabet{\mathbfsl}{OT1}{ppl}{b}{it} 

\newcommand{\vy}{\mathbfsl{y}}

\newcommand{\cC}{\mathcal{C}}
\newcommand{\cN}{\mathcal{N}}

\newcommand{\bR}{\mathbb{R}}
\newcommand{\bS}{\mathbb{S}}

\begin{document}

\title{Neural Network Decoders for Permutation \\ Codes Correcting Different Errors}

\author{\IEEEauthorblockN{Yeow Meng Chee
and Hui Zhang}\\
\IEEEauthorblockA{
Industrial Systems Engineering and Management, National University of Singapore, Singapore \\
Emails: ymchee@nus.edu.sg, isezhui@nus.edu.sg
}
}

\maketitle

\begin{abstract}
Permutation codes were extensively studied in order to correct different types of errors for the applications on power line communication and rank modulation for flash memory. In this paper, we introduce the neural network decoders for permutation codes to correct these errors with one-shot decoding, which treat the decoding as $n$ classification tasks for non-binary symbols for a code of length $n$. These are actually the first general decoders introduced to deal with any error type for these two applications. The performance of the decoders is evaluated by simulations with different error models.
\end{abstract} 


\section{Introduction}

\subsection{The Background}

The permutation code is a subset of the symmetric group, and was extensively studied because of potential applications on power line communication (PLC) and rank modulation (RM) for flash memory. Typically, the errors occurring in these applications include, but not limit to, insertion, deletion, substitution, adjacent transposition, and translocation. 

The application of permutation code on PLC channel is in $M$-ary FSK modulation scheme where symbols are modulated as sinusoidal waves with $M$ different frequencies \cite{Vinck2000,CCD2004,FVSB2005,SF2007,CP2012,HPS2015}. The noises prone to occur include additive background noise, impulse noise and permanent frequency disturbance, which can be considered as substitution errors on the permutation matrix of the corresponding codeword. In order to measure the PLC channel model with synchronization issue, it is natural to consider the mixture of these three types of errors and insertion/deletion errors. Nevertheless, the decoding of these mixed errors is a difficult task to achieve with classical decoding algorithms. 

For PLC channel without synchronization issue, decoding algorithms for permutation code were provided in \cite{HPS2015} and \cite{SF2007,CP2012} for codes obtained by composition of cosets of permutation groups and distance preserving mappings respectively. For PLC channel with synchronization issue, algorithms for correcting mixture of insertion/deletion/substitution errors for permutation codes were explored in \cite{CSF2008,CSF2009,HF2010,HFS2012,SSFT2012}. Their work mainly focused on some special cases, and no efficient general decoder is known yet. 
In their model, the output of the channel which suffers from the mixed errors is a variation of the original permutation instead of the corresponding permutation matrix, which is not necessary in the PLC channel model for regaining synchronization we consider in this paper. 

In RM scheme for flash memory, information is stored in the form of rankings of cell charges \cite{JMSB2008a,JSB2008b}. The translocation error, which is an extension of another well-studied error, the adjacent transposition error \cite{JSB2008b,BM2010,MBZ2013,BE2014,CV2014}, is caused by moving the rankings of one cell below a certain number of closest ranked cells. Interleaved codes correcting translocation errors equipped with certain Ulam distance were constructed in \cite{FSM2013}. However no efficient decoder for the general family of interleaved codes was known yet.

\subsection{The Neural Network Decoders}

Recently, a lot of research has been done on decoding with neural networks, we call them {\em neural network decoders}, see for example \cite{GCHB2017,LZJQZ2018,NBB2016,LG2017,NMLG2018,ZZL2020,LBM2019,LHP2018,LCHP2019,BHPSA2020,BCK2018,BRRB2020,MMAJ2019,CBCB2022}. Neural network decoders build upon supervised learning algorithms such as {\em multi-layer perceptron} (MLP), were proposed to decode codes as a classification task \cite{Haykin1998}. For binary codes, a single classification is replaced with $n$ binary classifications in \cite{GCHB2017,LZJQZ2018}, where $n$ is the length of code, and they showed that the bit error rate approaches the maximum a posteriori criterion decoding algorithms. 

Decoding permutation codes with non-binary symbols may induce more complexity compared to binary case, as we can see in next section that a codeword in a permutation code over $n$ symbols is transformed into square $n$ length bits for PLC application, which implies more decoding complexity even for codes of short length. Fortunately, in the applications of permutation codes, the length of codes is usually not quite large. For example, it was indicated in \cite{GLR2003} that, increasing the length of code utilized implies more critical constraints in terms of program and sensing accuracy. 

For linear codes, such as BCH codes, polar codes, Reed Solomon codes, deep learning succeeded on improving belief propagation decoding algorithms for large code length, see \cite{NBB2016,LG2017,NMLG2018,ZZL2020,LHP2018, LCHP2019,BHPSA2020,BCK2018,BRRB2020}.

\subsection{Our Contributions and Organization}

In this paper, we introduce low-latency one-shot neural network decoders for permutation codes for the applications of PLC and RM for flash memory. We use MLPs as our decoders, and let the output execute $n$ classification tasks as in \cite{GCHB2017,LZJQZ2018}, but for non-binary symbols, each corresponding to one coordinate of the codeword. Actually, these are the first general decoders introduced to deal with any error type for these two applications. Experiments show that the decoders can achieve good block error rate when the code has short length and small size, however the decoding ability decays when the code length and size increase.

The organization of the paper is as follows. In Section~\ref{sec:problem}, we formally state the error models of these applications, and also the permutation codes that we use in the paper. In Section~\ref{sec:neural}, we present the settings of the neural network decoders for permutation codes. In Section~\ref{sec:implement}, we show the performance by simulation, and Section~\ref{sec:conclusion} concludes the paper. 


\section{The Problem Statement}\label{sec:problem}

In this section, we formally introduce the error models of the two applications, and also provide some permutation codes that we use in simulation. 

\subsection{The Channel Model for PLC Channel}\label{subsection:plc}

In an $M$-ary FSK modulation scheme for power line communication (PLC), symbols are modulated as sinusoidal waves with $n$ different frequencies. In order to handle the noise, it was proposed in \cite{FVSB2005} that $n$ detected envelopes can be used in the decoding process. The $i$-th envelop detector for a frequency $f_i$, $i\in [1,n]\triangleq \{1,\dots,n\}$ is followed by a threshold $T_i$. For values above the threshold, it outputs a one, otherwise, a zero. Hence, we have $n$ outputs per transmitted symbol. A transmitted codeword of length $n$ thus leads to $n^2$ binary outputs, which are placed in a binary $n\times n$ matrix.

\begin{example}
\label{exp:plc_embed} Suppose a codeword $\pi=(1,2,4,3)$ is sent through the channel. Let $t_j$ represent the $j$-th time interval, for $j \in [1,4]$. If the codeword is received correctly, the output of the demodulator would be 
$$\begin{array}{ccccc}
f_1 & 1 & 0 & 0 & 0 \\
f_2 & 0 & 1 & 0 & 0 \\
f_3 & 0 & 0 & 0 & 1 \\
f_4 & 0 & 0 & 1 & 0 \\
 & t_1 & t_2 & t_3 & t_4
\end{array}$$
\end{example}


Three types of noises are prone to occur in the output matrix \cite{Vinck2000,FVSB2005}, namely, {\em additive background noise}: a one becomes a zero, or vice versa (with probability $p_{bg}$); {\em impulse noise}: a complete column is received as ones (with probability $p_{im}$); {\em permanent frequency disturbance}: a complete row is received as ones (with probability $p_{pfd}$). 

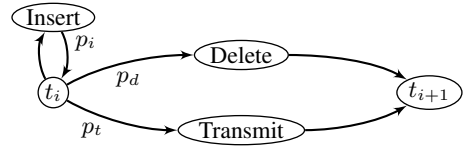
\begin{figure}[!t]
\centering
\begin{tikzpicture}
\tikzstyle{every node}=[font=\small]
\tikzset{mode/.style = {ellipse, draw=black, line width=0.5pt, inner sep=0.8pt}}
\tikzset{symbol/.style = {ellipse, draw, line width=0.5pt, align=center, inner sep = 0.8pt}}
\tikzset{arrow/.style = {->,> = latex',black,thick}}
	
\node[mode](insert) at (0,1.0) {Insert};
\node[mode](delete) at (2.5,0.5) {Delete};
\node[mode](transmit) at (2.5,-0.5) {Transmit};
\node[symbol](s1) at (0,0) {$t_i$};
\node[symbol](s2) at (5,0) {$t_{i+1}$};

\draw[arrow](s1) to [out=120, in=240, looseness=0.8] (insert);
\draw[arrow](insert) to [out=300, in=60, looseness=0.8] node [right, near start] {$p_i$} (s1);
\draw[arrow](s1) to [out=30, in=180, looseness=0.8] node [below] {$p_d$} (delete);
\draw[arrow](delete) to [out=0, in=150, looseness=0.8] (s2);
\draw[arrow](s1) to [out=330, in=180, looseness=0.8] node [below, near start] {$p_t$} (transmit);
\draw[arrow](transmit) to [out=0, in=210, looseness=0.8] (s2);
\end{tikzpicture}
\caption{The channel model for PLC. For each channel use, the ``insert'' state inserts a random symbol into the channel with probability $p_i$. With probability $p_d$, the next queued symbol is deleted, and with probability $p_t=1-p_i-p_d$, the next queued symbol is transmitted.}
\label{fig:channel_model}
\end{figure}

In this work, we also consider the PLC channel with synchronization issue, and employ the model in \cite{DM2001} (see Fig. \ref{fig:channel_model}). Suppose some symbols enter the queue to be transmitted over the channel. At each channel use, one of the three events occur: {\em (i)} with probability $p_i$, a random symbol is inserted and transmitted through the channel; {\em (ii)} with probability $p_d$, the next queued symbol is deleted; {\em (iii)} with probability $1-p_i-p_d$, the next queued symbol is transmitted through the channel, while also suffering from the above three types of noises from the channel. Here, the insertion/deletion errors correspond to a whole column inserted/missing in the output matrix. 

As in \cite{DM2001}, we assume $p_i=p_d$, and assume a maximum insertion length $\ell_{\max}$ for each time interval, and thus generate an output matrix of dimension $n\times (n+\ell_{\max}*(n+1))$ by filling zero columns at the end. In order to reduce storage waste with zero columns, we only take the first $c_{\max}$ columns in each matrix for a preset $c_{\max}$ in decoding phase. 

\begin{example}
\label{exp:plc} Take $\ell_{\max}=1$ and $c_{\max}=n+3$. Assume the codeword $(1,2,4,3)$ is sent through the channel, and the output matrix is
$$\begin{array}{cccccccc}
f_1 & 1 & 0 & 0 & 1 & 0 & 0 & 0 \\
f_2 & 1 & 1 & 1 & 1 & 0 & 0 & 0 \\
f_3 & 0 & 0 & 1 & 0 & 0 & 0 & 0 \\
f_4 & 0 & 1 & 0 & 0 & 0 & 0 & 0 \\
\end{array}$$
We can consider there exists a deletion when sending ``$2$'', a permanent frequency disturbance at frequency $f_2$, and an insertion at the last time slot. 
\end{example}

\subsection{The Error Model for RM Scheme}\label{subsection:rm}

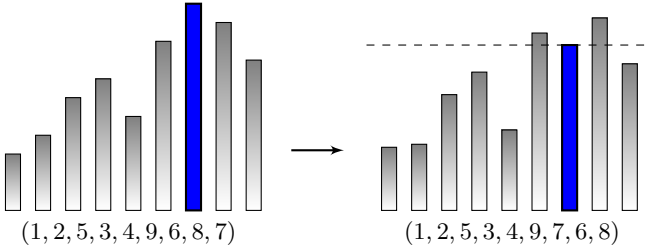
\begin{figure}[!t]
\centering
\begin{tikzpicture}
\tikzstyle{every node}=[font=\small]
\tikzset{arrow/.style = {->,> = latex',black,thick}}
\shadedraw [shading=axis] (0,0) rectangle (0.2,0.75);
\shadedraw [shading=axis] (0.4,0) rectangle (0.6,1.0);
\shadedraw [shading=axis] (0.8,0) rectangle (1.0,1.5);
\shadedraw [shading=axis] (1.2,0) rectangle (1.4,1.75);
\shadedraw [shading=axis] (1.6,0) rectangle (1.8,1.25);
\shadedraw [shading=axis] (2.0,0) rectangle (2.2,2.25);
\draw [fill=blue,thick] (2.4,0) rectangle (2.6,2.75);
\shadedraw [shading=axis] (2.8,0) rectangle (3.0,2.5);
\shadedraw [shading=axis] (3.2,0) rectangle (3.4,2.0);
\node[text width=40mm] at (2.2,-0.3) {$(1,2,5,3,4,9,6,8,7)$};
\shadedraw [shading=axis] (5.0,0) rectangle (5.2,0.84);
\shadedraw [shading=axis] (5.4,0) rectangle (5.6,0.88);
\shadedraw [shading=axis] (5.8,0) rectangle (6.0,1.54);
\shadedraw [shading=axis] (6.2,0) rectangle (6.4,1.84); 
\shadedraw [shading=axis] (6.6,0) rectangle (6.8,1.07);
\shadedraw [shading=axis] (7.0,0) rectangle (7.2,2.36);
\draw [fill=blue,thick] (7.4,0) rectangle (7.6,2.2);
\shadedraw [shading=axis] (7.8,0) rectangle (8.0,2.56);
\shadedraw [shading=axis] (8.2,0) rectangle (8.4,1.95);
\node[text width=40mm] at (7.3,-0.3) {$(1,2,5,3,4,9,7,6,8)$};
\draw [dashed] (4.80,2.2) -- (8.50,2.2); 
\draw[arrow] (3.8,0.8) to (4.5,0.8);
\end{tikzpicture}
\caption{The two figures show the original charge ranking and retrieved charge ranking suffering from Gaussian noises respectively.}
\label{fig:rank_model}
\end{figure}

In rank modulation (RM) for flash memory, the rank of a cell reflects the relative position of its own charge level, and the ranking of the $n$ cells induces a permutation (see Fig. \ref{fig:rank_model}). The data may be vulnerable to noises caused by potential cell over-injection, charge leakage, and read/write disturbance. The translocation errors were defined to characterize the noises \cite{JMSB2008a,JSB2008b,FSM2013}. For a permutation $\pi=(x_1,\dots,x_n)$, 
a {\em translocation} for distinct $i,j\in [1,n]$ is obtained by moving $x_i$ to the $j$-th position and shift elements between them, including $x_j$. Thus if $i<j$, the permutation $\pi$ after translocation $i$, $j$ is 
\begin{align*}
\cdots,x_{i-1},x_{i+1},\cdots,x_j,x_i,x_{j+1},\cdots
\end{align*}
and if $i>j$, it is
\begin{align*}
\cdots,x_{j-1},x_i,x_j,\cdots,x_{i-1},x_{i+1},\cdots.
\end{align*}

In order to characterize all the noises, we set our model as the analysis in \cite{FSM2013}. We assume that each cell charge suffers from a small Gaussian noise $n_1\sim\cN(0,\sigma_1^2)$ which may due to read/write disturbance and a low nearly uniform charge leakage rate, and with a small probability $p$, it also suffers from a possible large Gaussian noise $n_2\sim\cN(0,\sigma_2^2)$, for some $\sigma_2>\sigma_1$, which may caused by serious cell over-injection and high charge leakage rate. That means, suppose the charge of the $i$-th cell of the memory is $c_i$, and the charge retrieved is $c_i+n_1+n_2$. Apparently, the ranking of the charge levels of all the cells may be reordered. 

\begin{example}\label{exp:rm}
Assume $n=9$, $\sigma_1=0.2$, $p=0.001$, $\sigma_2=1.0$. In Fig.~\ref{fig:rank_model}, we suppose the original charges of the cells are $[1.5,2.0,3.0,3.5,2.5,4.5,5.5,5.0,4.0]$\footnote{Note that we take the charge voltage level arrangment similar as in \cite{GLR2003} for simulation in this paper.},
which corresponds to the permutation $(1,2,5,3,4,9,6,8,7)$, and the charges retrieved suffering from the Gaussian noises are $[1.68,1.76,3.08,3.68,2.14,4.72,4.40,5.12,3.90]$ corresponding to permutation $(1,2,5,3,4,9,7,6,8)$.
\end{example}

\subsection{Permutation Codes and Decoding Algorithms}\label{subsec:perm}



A permutation code is a subset of the symmetric group $\bS_n$, which consists of all permutations of $n$ symbols $1,2,\dots,n$.

The following Tenengolts’ single insertion/deletion correcting code is well known (see \cite{Tenengolts1984,Levenshtein1991}). 
For any $\pi=(x_1,\dots,x_n)\in\bS_n$, define the vector $\alpha(\pi)$ as
\begin{align*}
\alpha(\pi)_i=\begin{cases}
1, \text{ if $x_{i+1}\geq x_i$}. \\
0, \text{ otherwise}.
\end{cases}
\end{align*}
Define the code as the set 
\begin{align*}
\cC_n=\{\pi\in\bS_n:\sum_{i=1}^{n-1}i\alpha (\pi)_i\equiv 0\pmod{n}\}.
\end{align*}
However, this code may not be good at correcting substitution errors because of its low Hamming distance. We take the code $\cC_n^e$, which consists of only even permutations in $\cC_n$ and has minimum Hamming distance at least three. 

Permutation code with Hamming distance exactly characterizes PLC channel model without synchronization issue (see Appendix). Decoding algorithms were only provided in \cite{HPS2015} and \cite{SF2007,CP2012} for permutation codes obtained by composition of cosets of permutation groups and distance preserving mappings respectively in literature as far as we know. For comparison with the neural network decoders in this paper, we also provide a minimum distance (MD) decoding in Appendix. For PLC channel with synchronization issue, algorithms for correcting mixture of insertion/deletion/substitution errors for permutation codes were explored in \cite{CSF2008,CSF2009,HF2010,HFS2012,SSFT2012}. However their work mainly focused on some special cases. In particular, decoders of $\cC_n$ for the case a single error per codeword were provided in \cite{CSF2008,CSF2009}, and no general classical decoder is known yet.

For RM scheme, we take the interleaved code of size $((n/3)!/2)^3$ from \cite[Proposition 15]{FSM2013}, denoted as $\cC_n^{\rm IL}$, which was proved to correct a single translocation error, obtained by interleaving even permutations with $n/3$ symbols. Extensions into $t$-translocation error correcting codes were also provided in \cite{FSM2013}. However no efficient decoder for the general family of interleaved codes was known. 


\section{The Neural Network Decoders}\label{sec:neural}

In this section, we present the settings of the neural network decoders, that are employed to decode permutation codes. The readers may refer to \cite{Haykin1998} for more definitions.

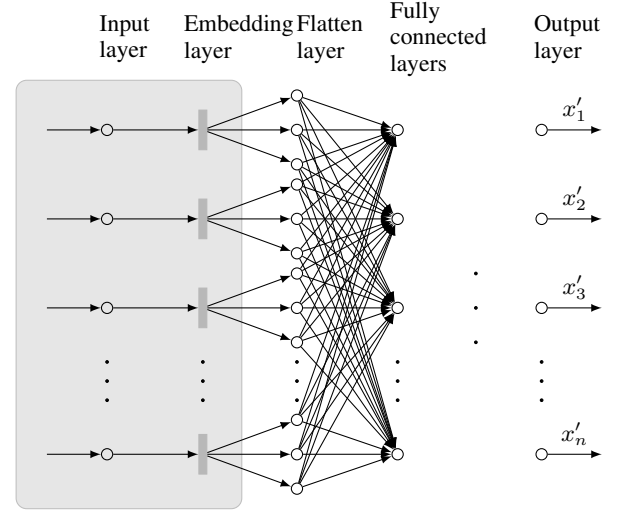
\begin{figure}
\centering
\begin{tikzpicture}[
plain/.style={
	draw=none,fill=none
},
rectangle/.style={
	fill=black,opacity=0.2,draw,shape=rectangle,minimum height=15pt
},
dot/.style={
	draw,shape=circle,minimum size=1.0pt,inner sep=0,fill=black
},
net/.style={
	matrix of nodes,
	nodes={
	draw,
	circle,
	inner sep=1.5pt
},
nodes in empty cells,
column sep=0.4cm,
row sep=3.0pt
},
>=latex
]
\tikzstyle{every node}=[font=\small]
\draw [draw=lightgray, fill=black!10, rounded corners] (-4.1,-3.5) rectangle (-1.1,2.2);
\matrix[net] (mat)
{
	|[plain]| \parbox{0.2cm}{Input\\layer} & |[plain]| \parbox{0.5cm}{Embedding\\layer} & |[plain]| \parbox{0.0cm}{Flatten\\layer} & |[plain]| \parbox{0.2cm}{Fully \\ connected\\layers} & |[plain]| & |[plain]| \parbox{0.2cm}{Output\\layer} \\
	|[plain]| & |[plain]| & \\
	& |[rectangle]| &  &  & |[plain]| & \\
	|[plain]| & |[plain]| &   \\
	|[plain]| & |[plain]| & & |[plain]| & |[plain]| \\
	& |[rectangle]| &  &  & |[plain]| & \\
	|[plain]| & |[plain]| &  & |[plain]| & |[plain]| \\
	|[plain]| & |[plain]| &  & |[plain]| & |[dot]| \\
	& |[rectangle]| &  &  & |[dot]| & \\
	|[plain]| & |[plain]| &  & |[plain]| & |[dot]| \\
	|[dot]| & |[dot]| & |[dot]| & |[dot]| & |[plain]| & |[dot]| \\
	|[dot]| & |[dot]| & |[dot]| & |[dot]| & |[plain]| & |[dot]| \\
	|[dot]| & |[dot]| & |[dot]| & |[dot]| & |[plain]| & |[dot]| \\
	|[plain]| & |[plain]| & \\
	& |[rectangle]| &  & & |[plain]| & \\
	|[plain]| & |[plain]| & \\
};
\foreach \ai/\mi in {3/1,6/2,9/3,15/n}
{
	\draw[<-] (mat-\ai-1) -- +(-0.8cm,0);
	\draw[->] (mat-\ai-1) -- (mat-\ai-2);
}

\foreach \ai in {2,3,4}
{
	\draw[->] (mat-3-2) -- (mat-\ai-3);
	\foreach \aii in {3,6,9,15}
		\draw[->] (mat-\ai-3) -- (mat-\aii-4);
}
\foreach \ai in {5,6,7}
{
	\draw[->] (mat-6-2) -- (mat-\ai-3);
	\foreach \aii in {3,6,9,15}
		\draw[->] (mat-\ai-3) -- (mat-\aii-4);
}
\foreach \ai in {8,9,10}
{
	\draw[->] (mat-9-2) -- (mat-\ai-3);
	\foreach \aii in {3,6,9,15}
		\draw[->] (mat-\ai-3) -- (mat-\aii-4);
}
\foreach \ai in {14,15,16}
{
	\draw[->] (mat-15-2) -- (mat-\ai-3);
	\foreach \aii in {3,6,9,15}
		\draw[->] (mat-\ai-3) -- (mat-\aii-4);
}
\draw[->] (mat-3-6) -- node[above] {$x_1'$} +(0.8cm,0);
\draw[->] (mat-6-6) -- node[above] {$x_2'$} +(0.8cm,0);
\draw[->] (mat-9-6) -- node[above] {$x_3'$} +(0.8cm,0);
\draw[->] (mat-15-6) -- node[above] {$x_n'$} +(0.8cm,0);
\end{tikzpicture}
\caption{The Multi-layer Perceptron Decoder}
\label{fig:MLP}
\end{figure}

\subsection{The Setting of Neural Network Decoders}

A {\em multi-layer perceptron} (MLP) is a class of feed-forward neural network, consisting three types of layers: {\em input layer}, {\em hidden layer}, and {\em output layer}. We describe explicitly the format of different layers for the two applications below. 

For RM model, the input layer is just the permutation retrieved from the cells that suffers from Gaussian noises, for example the permutation $(1,2,5,3,4,9,7,6,8)$ in Example~\ref{exp:rm}. We take the first hidden layer of the decoder as an {\em embedding layer}, namely a mapping $f:\{1,\dots,n\}\rightarrow \bR^*$, which maps each element of $1, \dots,n$ into a real vector of fixed length, and it is followed by a {\em flatten layer} that flats the input of this layer into a one-dimension vector, and several {\em fully connected} (or {\em dense}) layers that are connected to every neuron of its preceding layer, and an output layer (see Fig.~\ref{fig:MLP}). 

As for PLC, we just take the output matrix of the channel of dimension $n\times c_{\max}$ as given in Example~\ref{exp:plc} as the input layer, which is then followed by a flatten layer, several fully connected layers, and an output layer. 

In both cases, the output layer corresponds to the codeword to be retrieved, that was fed into the error models. Actually, the decoders execute $n$ classification tasks, where each of them corresponds to one coordinate of the codeword.  

\subsection{The Activation Functions and Output Layers}

Each fully connected layer performs an {\em activation function} $g(WX+b)$, which is taken to be rectified linear unit (ReLU) function $g(Z)_i=\max\{0,z_i\}$ for $Z=(z_1,z_2,\dots)$, where $X=(x_1,x_2,\dots)$ represents the vector received from the neurons in the previous layer. 

In the output layer, the $k$-th output performs a softmax function
\begin{align*}
g(Z_k)_i=e^{z_{k,i}}/(\sum_j e^{z_{k,j}})
\end{align*}
for $i,k\in [n]$, where $Z_k\triangleq (z_{k,1},\dots,z_{k,n})=W_kX+b_k$. Finally, $x_k'={\rm argmax}_i{g(Z_k)}_i$ is taken as the predicted symbol of the $k$-th coordinate of the codeword (see Fig.~\ref{fig:MLP}).

The ReLU function works to make the network approximate nonlinear functions \cite{HSW1989}, and softmax function normalizes the output to a probability distribution in each of the $n$ classification tasks.


%
%
%
%
%
%
%
%

\subsection{The Loss Function}

The loss function measures the difference between actual labels and the predicted output. Suppose $\pi=(x_1,\dots,x_n)$ is the codeword fed into the model. Let $\vy_k$ be the one-hot vector of $x_k$, which is a vector of length $n$ with all zero except a one at the $x_k$-th coordinate, and $\hat{\vy}_k$ be the output of the $k$-th softmax function, then the loss is defined as the sum of cross-entropy loss of the $n$ outputs, that is
\begin{align*}
Loss= -\sum_{k\in [n]}\sum_{i\in [n]}y_{k,i}\log(\hat{y}_{k,i})
\end{align*}
where $y_{k,i}, \hat{y}_{k,i}$ are the $i$-th components of $\vy_k$, $\hat{\vy}_k$ respectively.

\section{Implementation and Performance}\label{sec:implement}
 
In this section, we show the performance of neural network decoders introduced in the previous section by simulation. We take $\ell_{\max}=1$ and $c_{\max}=n+3$ or $n$ for PLC model with or without synchronization issue respectively. For RM model, we take the charge levels as $\{1.5+0.5i:i=[0,8]\}$ for $n=9$ and $\{1.5+0.4i:i=[0,11]\}$ for $n=12$.

\subsection{The Training/Test Sets and Hyper-Parameters}

The supervised learning algorithm produces an inferred function based on the labeled {\em training set}, and the {\em test set} is used for verification of the effectiveness of the training process. In the simulations, both the training and test sets are sets of data pairs composed of input and output of the neural network decoders as indicated in the previous section. 

In particular, for each point in Fig.~\ref{fig:comp_plc}--Fig.~\ref{fig:comp_rm}, we take a test set of size $10^6$, where each original codeword $\pi$ fed into the error models is randomly chosen from the whole codebook. In order to better visualize the decoding ability for different channel requirement, we train a model for each curve in the figures. For each curve, we let the training set has size $\delta M$ where $M$ is the size of code, and $\delta$ is the number of time that each codeword is fed into the error models in generating the training set pairs. For example, when $n=6$, for the curve with $p_i=p_d=0$ and $p_{im}=p_{pfd}=0.001$, we take the size of training set as $10^6M$, where each $10^5M$ of them corresponds to errors with the same $p_{im}$, $p_{pfd}$, $p_i$, $p_d$, and one $p_{bg}$. 

We let the MLP decoders all have three fully connected layers of the same size, and the embedding size is $n$ if necessary. In order to relief over-fitting, there is a dropout layer with rate $0.1$ before the output layer. The number $\delta$, size of dense layers and the total number of trainable parameters are listed in Table~\ref{tab:par_num}. 
Notice that, in the first four rows of the last column, the values denote the trainable parameters for PLC model with (or without) synchronization issue respectively.

We implement with TensorFlow library. In the training process, we take batch size as $200$, and use Adam algorithm for optimization with a default learning rate $0.001$. In order to fully utilize the data in training set, we choose the number of epochs larger than one in the way that the decoders achieve a higher BLER on the test set, which indicates the rounds that entire training set is passed in training process. 


\begin{table}[t]
\begin{center}
\begin{tabular}{c|c|c|c|c}
\hline
code & size & $\delta$ & dense layers & $\#$parameters \\
\hline
$\cC_6^e$ & $56$ & $10^6$ & 128 & $44,708$ ($42,404$) \\
$\cC_7^e$ & $360$ & $10^5$ & 128 & $48,433$ ($45,745$) \\
$\cC_8^e$ & $2544$ & $10^4$ & 128 & $52,672$ ($49,600$) \\
$\cC_8^e$ & $2544$ & $10^4$ & 256 & $170,816$ ($164,672$) \\
\hline
$\cC_9^{\rm IL}$ & $27$ & $10^6$ & 64 & $18,914$ \\
$\cC_{12}^{\rm IL}$ & $1728$ & $10^4$ & 64 & $27,104$ \\
\hline
\end{tabular}  
\end{center}
\caption{The Parameters of MLP Decoders.}
\label{tab:par_num}
\end{table}


\subsection{The Performance and Analysis}

The performance of our schemes is measured by the block error rate (BLER) decoding on test set. 

\begin{figure}[t!]
\includegraphics[width=10cm, center]{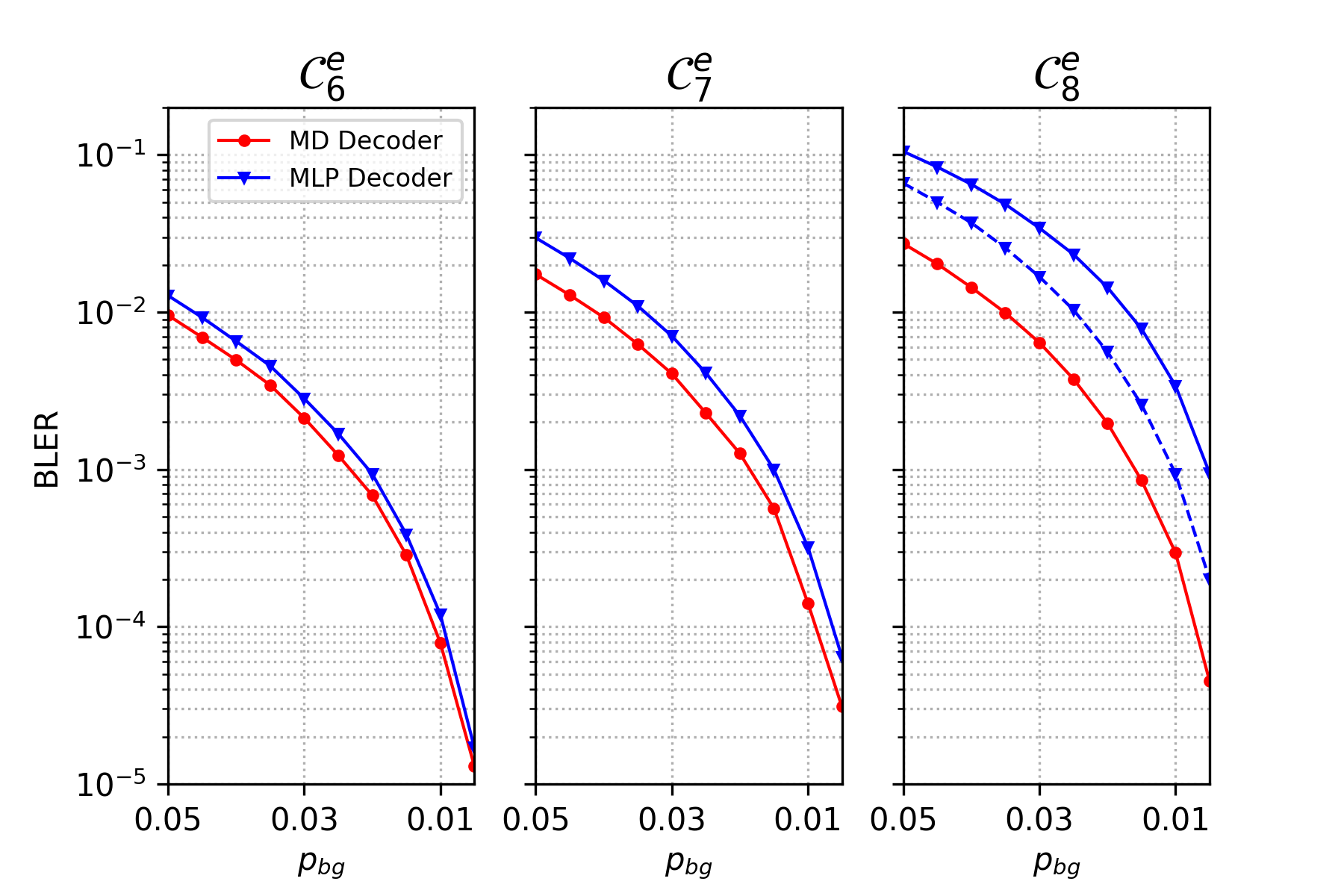}
\caption{The Performance of Decoders for PLC Channel without Synchronization Issue: The solid line and dashed line are for MLP decoders with hidden layers size $128$ and $256$ respectively. Here, $p_{im}=p_{pfd}=0.001$.}
\label{fig:comp_plc}
\end{figure}

In Fig.~\ref{fig:comp_plc}, we consider the PLC channel without synchronization issue. We take the test sets with $p_{im}=p_{pfd}=0.001$, and the background noises in the set $\{0.005i:i\in[1,10]\}$. The MD decoder is implemented for comparison. We can see that when $n=6$, the MLP decoder approaches the MD decoder, however when $n$ increases, the gap also increases. For $n=8$, an MLP decoder with dense layer size $256$ performs better compared to the one with size $128$, and more trainable parameters lead to better decoding ability in this case.  

\begin{figure}[t!]
\includegraphics[width=10cm, center]{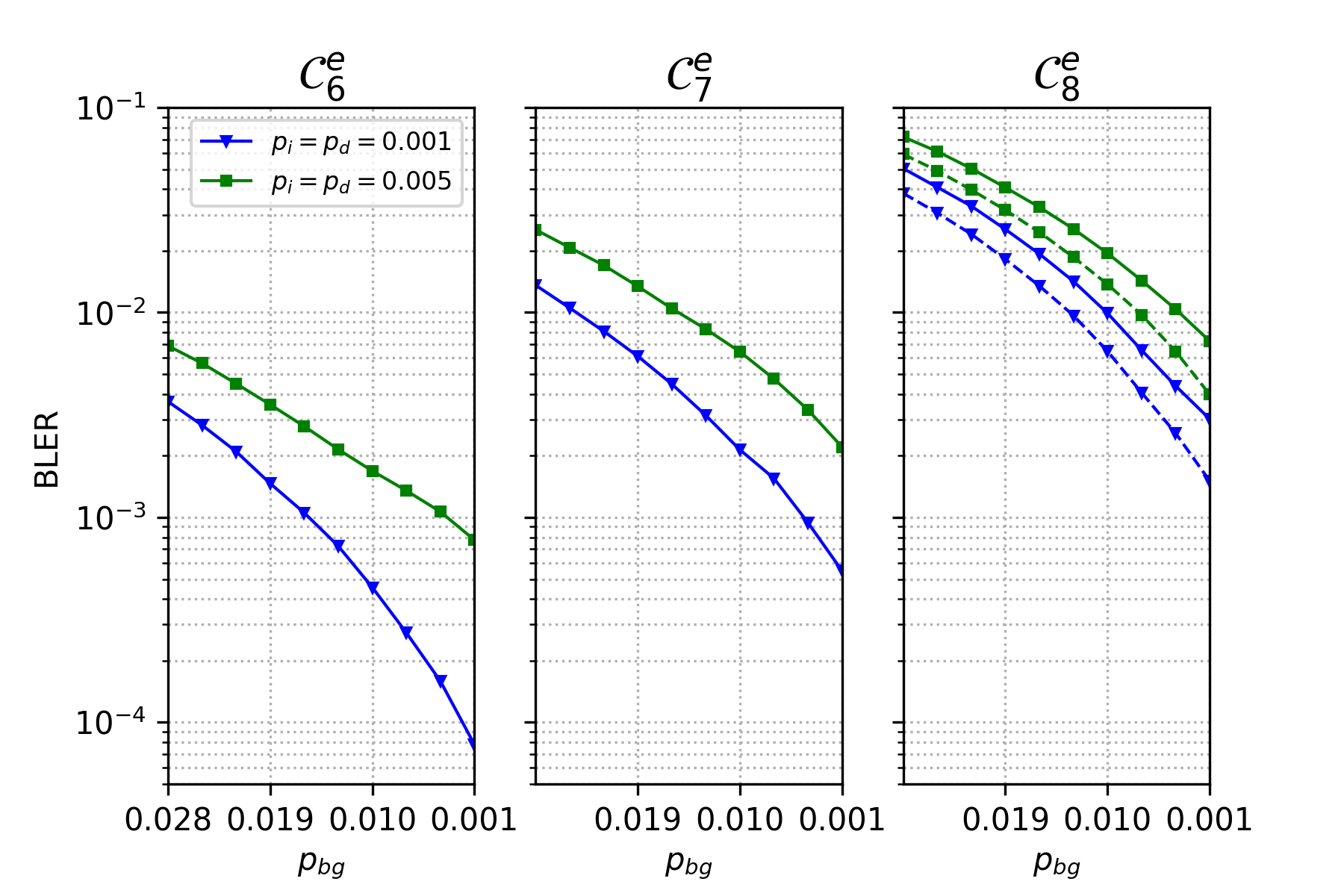}
\caption{The Performance of Decoders for PLC Channel with Synchronization Issue: The solid line and dashed line are for MLP decoders with hidden layers size $128$ and $256$ respectively. Here, $p_{im}=p_{pfd}=0.001$.}
\label{fig:comp_plc1}
\end{figure}

In Fig.~\ref{fig:comp_plc1}, we consider the PLC channel with synchronization issue. We take the test sets with $p_{im}=p_{pfd}=0.001$, and the background noises in the set $\{0.001+0.003i:i\in[0,9]\}$, for each $p_i=p_d\in \{0.001,0.005\}$. We can see that when $n=6$, the BLER can reach below $10^{-4}$ for $p_i=p_d=p_{bg}=0.001$. However, when $n$ increases, the decoding ability seems decay faster than in Fig.~\ref{fig:comp_plc}. For $n=8$, the decoder with dense layer size $256$ also performs better in this case. 

\begin{figure}[t]
\includegraphics[width=10cm, center]{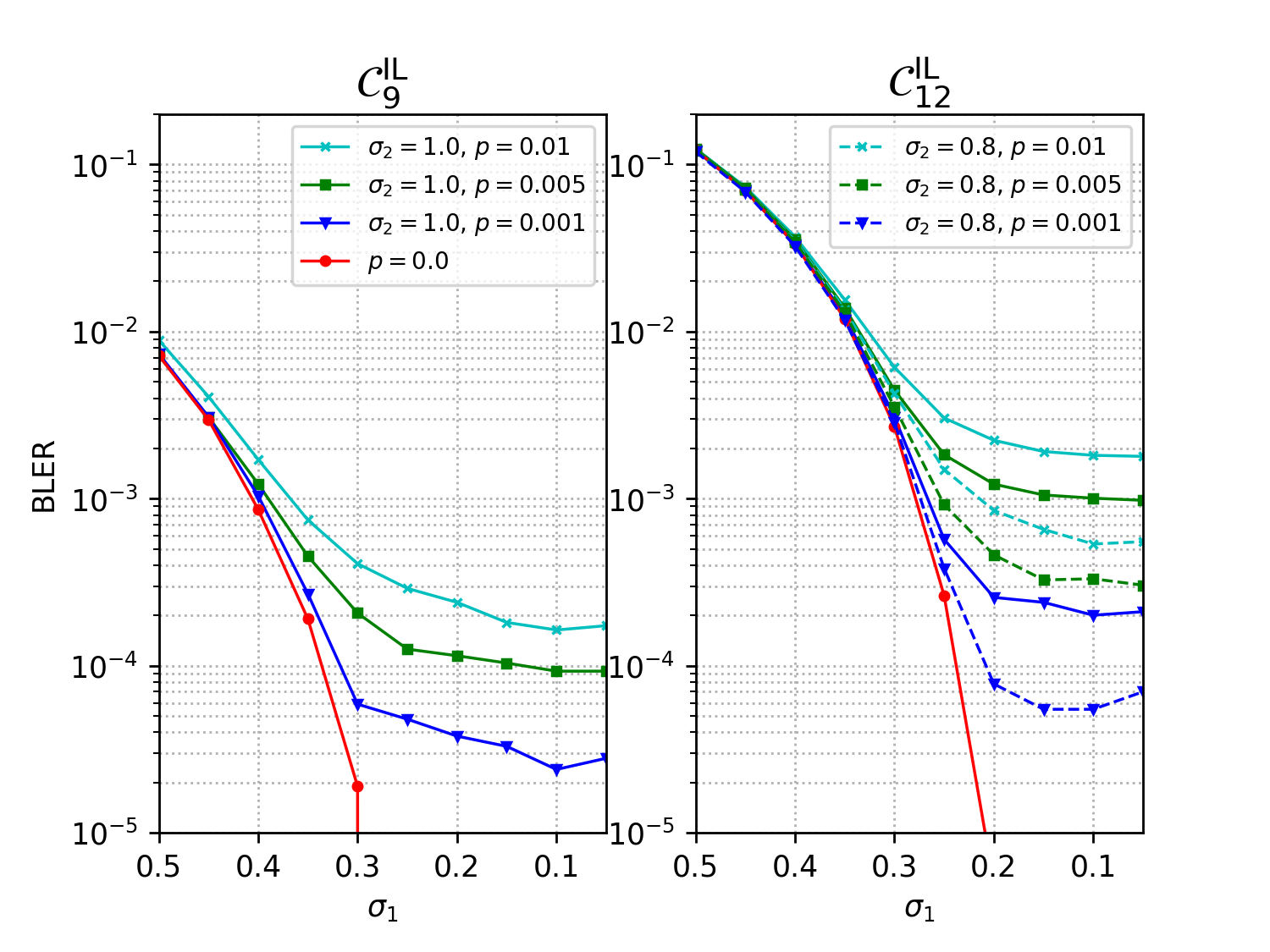}
\caption{The Performance of MLP Decoders for RM Scheme.}
\label{fig:comp_rm}
\end{figure}

In Fig.~\ref{fig:comp_rm}, we consider the MLP decoders for RM scheme for $n\in \{9,12\}$. We take the test sets with $\sigma_1$ in the set $\{0.05i:i\in[1,10]\}$. When $p=0$, that is the scheme only suffers from the small disturbance, both the codes can reach zero BLER when $\sigma_1$ is small. When $\sigma_2$ is twice the gap of two adjacent charge levels, the BLER for $n=9$ reaches below $10^{-4}$ when $p\in \{0.001,0.005\}$ for small $\sigma_1$, and it also holds for $n=12$, when $p=0.001$. However, when taking fixed $\sigma_2=1$, the code with length $12$ has less reliability.

In our decoders, the decoding is treated as $n$ classification tasks. In essence, in training process when labeled data is fed into the neural network, the local features are memorized by trainable parameters, and in decoding phase, the network infers the label of data by recognizing the features. Therefore, compared to the classical decoding algorithms, the neural network decoder may provide more flexibility. 

\section{Conclusion}\label{sec:conclusion}

In this paper, we introduced neural network decoders for one-shot decoding of permutation codes correcting errors for PLC and RM for flash memory, which are the first general decoders introduced to correct any error types. Experiments show that the decoders perform well on correcting the errors, in particular for codes of small length and size. The decoding ability decays when the length and size of codes increase. It will be interesting to explore in future work whether it is possible to increase the decoding ability by further increasing the complexity of neural network decoders. 

\section*{Acknowledgment}

The authors would like to thank Zekun Tong and Xinke Li for assistance on programming. 

\section*{Appendix}\label{sec:appendix}

The connection between permutation codes with Hamming distance and PLC channel model was implied by not given explicitly in literature. Here, we provide a proof. For a permutation $\pi=(x_1,\dots,x_n)$, we let $Y^{\pi}$ be the $n\times n$ matrix by assigning one in the $x_i$-th row in the $i$-th column for $i\in[1,n]$ and zero in all other positions, and let $Y^{\cC}=\{Y^\pi:\pi\in\cC\}$. Let $\pi, \sigma\in \cC$, and the Hamming distances $d_H(\pi, \sigma)$ and $d_H(Y^{\pi}, Y^{\sigma})$ are defined as the number of positions that the two components differ in. 

\begin{proposition}
A permutation code can correct any combination of $e_1$ background noise, $e_2$ impulse noise, and $e_3$ permanent frequency disturbance with $e_1+e_2+e_3\leq d-1$ if and only if it has minimum Hamming distance $d$.
\end{proposition}
\begin{IEEEproof}
For any $\pi, \sigma\in \cC$, since $d_H(\pi,\sigma)\geq d$, we have  $d_H(Y^{\pi},Y^{\sigma})\geq 2d$. In the output matrix $M$ suffering from noise, firstly we can locate the rows $I$ and columns $J$ that the impulse noise and permanent frequency disturbance occur, since they are all-one vectors. Then we can consider these rows and columns as erasures. Omitting the chosen rows and columns in all the matrices in $Y^{\cC}$, the resulting set of matrices still has minimum Hamming distance at least $2d-2e_2-2e_3$, and can correct up to $\lfloor(2d-2e_2-2e_3-1)/2\rfloor=d-1-e_2-e_3$ background noise. The reverse is obvious, since if the permutation code has minimum Hamming distance $d-1$, then some $d-1$ background noise (or impulse noise, permanent frequency disturbance) may ruin the decoding ability of $\cC$.
\end{IEEEproof}

We denote the matrix $Y$ by omitting the rows in $I$ and columns in $J$ as $Y\backslash\{I,J\}$. For a received matrix $M$ of the channel output, we can decode $M$ as the codeword
\begin{align}
\pi^\star = {\sf argmin}_{\pi\in \cC} d_H(Y^\pi\backslash\{I,J\},M\backslash\{I,J\}) \label{mdd_eq1}
\end{align}
Since the decoding by \eqref{mdd_eq1} is time costly, in Fig.~\ref{fig:comp_plc} we implement the minimum distance (MD) decoding by the following
\begin{align}
\pi^\star = {\sf argmin}_{\pi\in \cC} d_H(Y^\pi,M) \label{mdd_eq2}
\end{align}
When the probabilities of impulse noise, and permanent frequency disturbance are low, that is, mostly only one of them occurs, the decoding with \eqref{mdd_eq2} that differs from decoding with \eqref{mdd_eq1} will be rare. 


\end{document}